\RequirePackage{fix-cm}
\documentclass[twocolumn,epjc3]{svjour3}  % EPJC

\smartqed  % flush right qed marks, e.g. at end of proof

\usepackage{graphicx}  % needed for figures
\usepackage{dcolumn}   % needed for some tables
\usepackage{bm}        % for math
\usepackage{amssymb}   % for math
\usepackage{sidecap}   % sidecaptions
\usepackage{subfigure}
\usepackage{amsmath}
\usepackage{pdfpages} % for authorlist
\usepackage{multirow}
\usepackage{xspace}
\usepackage{booktabs}
\usepackage[normalem]{ulem}
\usepackage[numbers,sort&compress]{natbib}
\usepackage{epstopdf} %allows to use pdftex
\usepackage{placeins}

\usepackage{hyperref}

\usepackage{preprintcover}  % load package
\PreprintCoverPaperTitle{Measurement of the $t\bar{t}$ production cross
  section in the tau+jets channel using the ATLAS detector}  % paper title 
\PreprintIdNumber{CERN-PH-EP-2012-328}  % CERN preprint number
\PreprintCoverAbstract{A measurement of the top quark pair
  production cross section in the final state with a hadronically decaying
  tau lepton and jets is presented. The analysis is based on proton--proton
  collision data
  recorded by the ATLAS experiment at the LHC, with a centre-of-mass energy
  of 7 TeV. The data sample corresponds to an integrated luminosity of
  1.67~\ifb{}. The cross section is measured to be $\sigma_{t\bar{t}}$~=~194
  $\pm$ 18 (stat.) $\pm$ 46 (syst.)~pb and is in agreement with other
  measurements and with the Standard Model prediction.} % abstract
\PreprintJournalName{EPJC}  % journal name
\usepackage{atlasphysics} 
\usepackage{subfigure}

\journalname{Eur.\ Phys.\ J.\ C}
\input{commands}

\begin{document}

\title{Measurement of the $t\bar{t}$ production cross section
  in the tau+jets channel using the ATLAS detector} 

\author{The ATLAS Collaboration}
\institute{CERN, 1211 Geneva 23, Switzerland}
\date{\today}

\maketitle % for EPJC / RevTex

\begin{abstract}

  A measurement of the top quark pair
  production cross section in the final state with a hadronically decaying
  tau lepton and jets is presented. The analysis is based on proton--proton collision data
  recorded by the ATLAS experiment at the LHC, with a centre-of-mass energy
  of 7 TeV. The data sample corresponds to an integrated luminosity of
  1.67~\ifb{}. The cross section is measured to be $\sigma_{t\bar{t}}$ = 194
  $\pm$ 18 (stat.) $\pm$ 46 (syst.)~pb and is in agreement with other
  measurements and with the Standard Model prediction.

\end{abstract}

% -*- mode: LaTeX; mode: auto-fill; fill-column: 75 -*-

\section{Introduction}

Top quark pairs (\ttbar) are produced in abundance at the Large Hadron
Collider (LHC) due to the high centre-of-mass energy of 7 TeV.  The large
sample of \ttbar events collected with the ATLAS detector makes it possible
to study experimentally challenging decay channels and topologies.  This
letter describes a measurement of the \ttbar production cross
section. The final state studied here consists of a hadronically decaying tau
lepton (\tauhad) and jets, corresponding to the $t\bar{t} \rightarrow
[b\tauhad   \nu_{\tau}][bqq]$ decay, where $b$ and $q$ are used to denote
$b$-quarks and lighter quarks, respectively.  Such an event topology with a
hadronically decaying tau lepton corresponds to approximately 10\% of all
\ttbar decays \cite{PDG}.

A \ttbar cross-section measurement in the final state with tau leptons makes
it possible to probe flavour-dependent effects in top quark decays. It is
also relevant to searches for processes beyond the Standard
Model, where \ttbar events with tau leptons in the final state are a
dominant background.  This measurement is particularly important for hypothetical
charged Higgs boson production
\cite{ChargedHiggs1,ChargedHiggs2,ChargedHiggs3,ChargedHiggsAtlas} in top
quark decays, where the existence of a charged Higgs boson would lead to an
enhancement in the cross section for the considered \ttbar final state. The measurement presented here is
complementary to the previously published tau + lepton (electron or muon)
channel measurement \cite{taulep}. The most recent cross-section
measurements of the tau + jets decay channel have been performed by the CDF
and D0 collaborations in proton--antiproton collisions at $\sqrt{s}$ =
1.96~\TeV{}~\cite{PhysRevD.82.071102,CDF}. This is the first measurement
reported in this specific channel at the LHC. 

In this analysis, events with at least five jets are selected, where two of
the jets are identified as having originated from $b$-quarks.  After
identifying the two jets likely to come from the hadronic decay of one of
the top quarks, one of the remaining jets is selected as the \tauhad
candidate from the other top quark. The \tauhad contribution is separated
from quark- or gluon-initiated jets with a one-dimensional fit to the
distribution of the number of tracks associated with the \tauhad
candidate. Since the \tauhad decays preferentially to one or three charged
particles (and other neutral decay products), this variable provides good
separation between hadronically decaying tau leptons and jets, as the
latter typically produce a large number of charged particles. The main
backgrounds to the \ttbar signal are multijet events, \ttbar events
with a different final state or signal events where the wrong jet is chosen
as the \tauhad candidate.  A small contribution from single-top and
$W$+jets events is also present. The distributions for the backgrounds used
in the fit are obtained with data-driven methods.

% -*- mode: LaTeX; mode: auto-fill; fill-column: 75 -*-
\section{The ATLAS detector}
\label{sec:ATLAS}

\FloatBarrier
The ATLAS detector~\cite{Aad:2008zzm} is a multipurpose particle physics
detector with a forward-backward symmetric cylindrical geometry and a near-4$\pi$ coverage in solid angle\footnote{ATLAS uses a right-handed
  coordinate system with its origin at the nominal interaction point (IP)
  in the centre of the detector and the $z$-axis along the beam pipe. The
  $x$-axis points from the IP to the centre of the LHC ring, and the $y$-axis points upward. Cylindrical coordinates $(r,\phi)$ are used in the
  transverse ($x$, $y$) plane, $\phi$ being the azimuthal angle around the beam
  pipe. The pseudorapidity is defined in terms of the polar angle $\theta$
  as $\eta=-\ln\tan(\theta/2)$.  The variable $\DeltaR$ is used to evaluate
  the distance between objects, and is defined as $\DeltaR =
  \sqrt{(\Delta\phi)^2 + (\Delta\eta)^2}$.}. The inner tracking system
covers the pseudorapidity range \abseta $<$ 2.5, and consists of a silicon
pixel detector, a silicon microstrip detector (SCT), and, for \abseta $<$
2.0, a transition radiation tracker. The inner detector is surrounded by a
thin superconducting solenoid providing a 2 T magnetic field along the beam
direction. A high-granularity liquid-argon sampling electromagnetic
calorimeter covers the region \abseta $<$ 3.2. An iron/scintillator tile
hadronic calorimeter provides coverage in the range \abseta $<$ 1.7. The
end-cap and forward regions, spanning 1.5 $<$ \abseta $<$ 4.9, are
instrumented with liquid-argon calorimeters for both electromagnetic and
hadronic measurements. The muon spectrometer surrounds the calorimeters. It
consists of three large air-core superconducting toroid systems and
separate trigger and high-precision tracking chambers providing accurate
muon tracking for \abseta $<$ 2.7.

% -*- mode: LaTeX; mode: auto-fill; fill-column: 75 -*-

\section{Data and simulation samples}
\label{sec:sample}

The data used in this analysis were collected during the first half of the
2011 data-taking period and correspond to a total integrated luminosity of
${\cal L} = 1.67$~\ifb.  The data sample was selected with a $b$-jet
trigger that required at least four jets identified with \abseta $<$ 3.2
and a transverse energy (\et{}) above 10~\GeV{}. Two of these jets were
required to be identified as $b$-jets using a dedicated high-level-trigger
$b$-tagging algorithm \cite{ATLASTriggerPerf}.  This trigger was enabled
for only part of the 2011 data-taking period and is therefore the
limiting factor in determining the integrated luminosity of the dataset
used.

The selection efficiency for the $\ttbar\to\tauhad$ + jets
signal is derived from Monte Carlo (MC) simulations. The {\sc MC@NLO}
v4.01~\cite{SAMPLES-MCNLO} generator, with the parton distribution function
(PDF) set {\sc CT10}~\cite{CT10}, is used for the \ttbar signal. The
theoretical prediction of the \ttbar cross section for proton--proton
collisions at a centre-of-mass energy of $\sqrt{s}$ = 7 \tev~is
$\sigma_{\ttbar}$ = $167^{+17}_{-18}$~pb for a top quark mass of
172.5~\GeV. It has been calculated at approximate next-to-next-to-leading
order (NNLO) in QCD with {\sc Hathor 1.2}~\cite{Aliev11} using the {\sc
  MSTW2008} 90\% confidence level NNLO PDF sets~\cite{Martin09}, incorporating
PDF and $\alpha_S$ uncertainties according to the {\sc MSTW}
prescription~\cite{MSTW09}, and cross-checked with the next-to-leading-order
+ next-to-next-to-leading-log calculation of Cacciari et al.~\cite{Cacciari} as implemented in {\sc
  Top++} 1.0~\cite{Czakon:2011xx}. Tau lepton decays are modelled with {\sc
  Tauola}~\cite{TAUOLA}. Samples of simulated events are also used to estimate the small
contributions from $W$+jets, $Z$+jets, single-top-quark and diboson events, as
described in Ref.~\cite{ATLAStopxs}. The generated events were processed
through the full ATLAS detector simulation using {\sc
  Geant4}~\cite{AGO-0301,SimulationPaper}, followed by the trigger and
offline reconstruction. The distribution of the number of pile-up events
(i.e. collisions in the same, or nearby, bunch crossing as the hard-scattering event)
is adjusted to match the scattering multiplicity measured in the data. 

% -*- mode: LaTeX; mode: auto-fill; fill-column: 75 -*-

\section{Event selection}
\label{sec:obj}

Jets are reconstructed from clusters of calorimeter
\linebreak cells~\cite{TopoClusters} using the anti-$k_{t}$
algorithm~\cite{Cacciari:2005hq,Cacciari:2008gp} with a radius parameter
$R=0.4$. The jets are calibrated using transverse momentum- and
$\eta$-dependent corrections obtained from simulation and validated with collision data~\cite{jetsys}.
Candidate events are required to contain at least five jets with a transverse
momentum (\pt) larger than 20~\GeV{} and \abseta $<$ 2.5.

The identification of jets originating from $b$-quarks is performed using
algorithms that combine secondary vertex properties and track
impact parameters~\cite{Btagging}. The algorithm identifies $b$-jets with
an average efficiency of 60\% and provides a light-quark jet rejection
factor of about 340 in \ttbar topologies. The likelihood of misidentifying
a \tauhad as a $b$-jet in a \ttbar event is approximately 5\%. The two
jets with the highest $b$-tag probability are chosen as the event
$b$-candidates; events with fewer than two $b$-jets are rejected. 

The magnitude of the missing transverse energy \linebreak (\met{}) is reconstructed from 
energy clusters in the calorimeters.  The calibration of each cluster
depends on the type of physical object associated with the cluster. The
transverse momentum of muons in the event is also taken into account.  The \met{}
significance (\metsig) is defined as $\met / (0.5~[\sqrt{\mathrm{GeV}}]
\cdot\sqrt{\sumet})$, where \sumet is the scalar sum of the transverse
momentum of all objects.  Using a \metsig requirement instead of a direct
\met requirement allows the rejection of multijet events where the \met{}
arises from energy resolution effects, while still retaining high
efficiency for signal events with \met{} coming from particles which do not
interact with the detector~\cite{ATLASW}. Candidate events are required to
have \metsig $>$ 8. 

Events containing a reconstructed electron or
\linebreak muon~\cite{Aad:2011mk,muonPerformance} with \pt $>$ 15~\GeV{} and \abseta
$<$ 2.5 are vetoed to reduce the background due to events containing $W$
bosons that decay to electrons or muons and to avoid overlap with other \ttbar
cross-section measurements.

\label{sec:sel}

In each event, a single \tauhad candidate is selected from the
reconstructed jets using the following procedure. First, the reconstruction
of the hadronically decaying top quark is attempted by selecting the three
jets (including exactly one of the two $b$-candidates) which, when their
four-momenta are added vectorially, give the highest \pt sum. The remaining
jet with the highest \pt{}, excluding the remaining $b$-candidate, is
selected as the \tauhad candidate. Events where the \tauhad candidate \pt
is below 40 GeV are rejected.

\label{sec:signal}

The main contributions to the selected \tauhad candidates in the signal
region come from the signal (\tauhad from \ttbar events), electrons from
\ttbar events and misidentified jets from \ttbar, single-top-quark production, $W$+jets and
multijet events. The contributions from $Z/\gamma^{\star}$+jets and diboson
processes are negligible.

% -*- mode: LaTeX; mode: auto-fill; fill-column: 75 -*-

\section{Data analysis}
\label{sec:fitvariable}

The majority of \tauhad decays are characterised by the presence of one or
three charged hadrons in the final state, which can be reconstructed as
charged particle tracks in the inner detector. The number of tracks (\ntrack{})
originating from the interaction point associated with a \tauhad candidate is
used to separate the \tauhad contribution from the misidentified jet background. 

\begin{figure}[htb]
  \centering
  \includegraphics[width=1.0\linewidth]{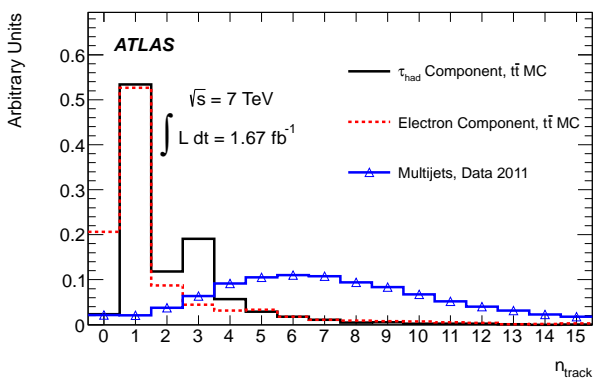}
  \caption{Distribution of \ntrack{} for \tauhad{} from MC \ttbar{} events
    (solid black line), electrons from MC \ttbar{} events (dashed red line), and for jets from multijet events from data (blue
    triangles).  The multijet event selection uses a \metsig sideband
    region as described in Sect.~\ref{sec:templates}. All distributions are normalised to unity.}
  \label{fig:ntrack}
\end{figure}

All selected tracks with \pt $>$ 1~\GeV{} located in a core region spanning
\DeltaR $<$ 0.2 around the jet axis are counted.  To increase the
discriminating power, tracks in the outer cone 0.2 $<$ \DeltaR $<$ 0.6 are
also counted, using a variable \pt{} requirement that is dependent on both the \DeltaR of
the outer track and the \pt{} of the core tracks.  This variable \pt{} requirement
is designed to reduce the contribution from pile-up and underlying event
tracks, and is explained in Ref.~\cite{Count}.  The separation
power of the \ntrack{} variable is illustrated in Fig.~\ref{fig:ntrack}
where a comparison of the \ntrack distribution is shown for \tauhad,
electrons and misidentified jets from multijet events.

\label{sec:templates}

To extract the signal from the \ntrack{} distribution, the data sample is
fitted with three probability density functions (templates):  a
\tauele{} template, a \qcd{} template and a \comb{} template. The \tauhad
component from \ttbar events constitutes the signal in the event
sample. Real electrons from \ttbar events (either prompt or from leptonic
tau decays) which failed to be rejected by the electron veto also contribute
significantly to the event sample. The electron and \tauhad templates are
combined into a single \tauele template to ensure a stable fit, using MC
predictions to determine their relative contributions. The \tauele
template is obtained from simulated \ttbar events. The small expected
contributions to the real \tauele component of the fit from single-top-quark and
$W$+jets events do not change the shape of the template.

The remaining significant contributions come from misidentified jets, and
are separated into two templates. The \qcd template describes the QCD multijet
processes which are dominated by gluon-initiated jets, and the \comb template
describes the remaining processes (\ttbar, single-top quark and $W$+jets) that
are enriched in quark-initiated jets.

The \qcd{} template is determined using a sideband region where the
\metsig{} requirement is changed to 3 $<$ \metsig $<$ 4.  This selection
greatly enhances the contribution from multijet events, reducing other
contributions (e.g. from \ttbar events) to less than 1\%.  The regions
defined by the selection 2 $<$ \metsig $<$ 3 and 4 $<$ \metsig $<$ 5 are
also used to study any correlations between the \metsig criteria and
the \ntrack distribution.

The \comb template is obtained from a \ttbar control sample where the \tauhad
candidate is replaced by a muon candidate. The reconstructed
muon~\cite{muonPerformance} is required to have \pt $>$ 20~\GeV, \abseta
$<$ 2.5 and no jet within a distance \DeltaR $=$ 0.4.  The requirement on the number
of non-$b$-tagged jets is changed from three to two as the jet
corresponding to the \tauhad is now replaced by a muon.  The other
selection requirements are the same as for the signal region. This isolates
\ttbar events with very high purity; the contribution from backgrounds is
estimated from MC predictions to be at the 5\% level, and consists mainly
of single-top-quark and $W$+jets events. The
two highest-\pt{} jets that are not identified as $b$-jet candidates are
selected as \tauhad candidates. The template is corrected using MC
simulations for differences in the transverse momentum distribution between the signal region and the
control sample, and for the expected contribution to the control sample
from \ttbar dilepton events ($\ttbar\rightarrow \mu + \tauhad + X$,
$\ttbar\rightarrow \mu + e + X$).

% -*- mode: LaTeX; mode: auto-fill; fill-column: 75 -*-

\section{Results}
\label{sec:fitResult}

An extended binned-likelihood fit is used to extract the different
contributions from the \ntrack distribution. To improve the fit stability,
a soft constraint is applied to the ratio of \comb events to \tauele
events, which are dominated by the same process (\ttbar events).  The
constraint, based on MC predictions, is a Gaussian with a width of 19\% of
its central value.  This width was estimated based on studies of the associated systematic
uncertainties using the same methodology as described in Sect.~\ref{sec:systematics}.  The statistical uncertainties on the fit parameters are
calculated using the shape of the fit likelihood. The systematic
uncertainties on the shapes of the templates are propagated using a
pseudo-experiment approach, taking into account the bin-by-bin
correlations. This yields a final number of \tauele events of 270 $\pm$ 24
(stat.) $\pm$ 11 (syst.).

\begin{figure}[htb]
  \centering
    \includegraphics[width=1.0\linewidth]{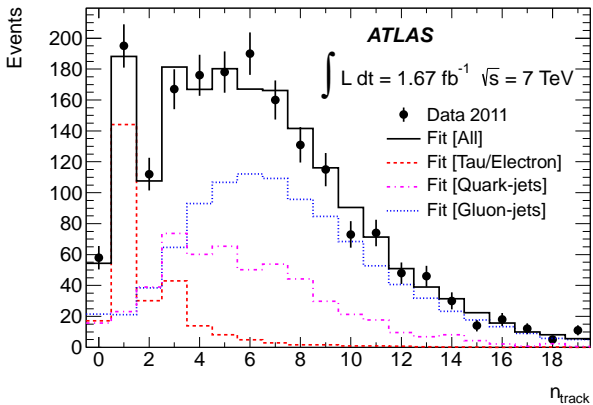}
    \caption{ The \ntrack distribution for \tauhad candidates after all
      selection cuts.  The black points correspond to data, while the
      solid black line is the result of the fit. The red (dashed), blue (dotted) and
      magenta (dash-dotted) histograms show the fitted contributions from the \tauele signal, and
      the \qcd and \comb backgrounds, respectively.}
  \label{fig:fit}
\end{figure}

The fit results are shown in Fig.~\ref{fig:fit}.  A comparison between the
fit results, and the expected event yields from the MC predictions is
presented in Table~\ref{tab:pred}.  The numbers are in good agreement.

\begin{table}[hbtp]
  \centering
  \begin{tabular}{|c|c|}

    \hline
    Source & Number of events \\
    \hline
    \multicolumn{2}{|c|}{\tauele} \\
    \hline
    \ttbar (\tauhad) & 170 $\pm$ 40 \\
    \ttbar (electrons) & 47 $\pm$ 11\\
    Single top & 12 $\pm$ 2\\
    $W$+jets & 9 $\pm$ 5\\
    \hline
    Total expected & 240 $\pm$ 50 \\
    \hline
    Fit result & 270 $\pm$ 24 (stat.) $\pm$ 11 (syst.) \\

    \hline
    \multicolumn{2}{|c|}{\comb} \\
    \hline
    \ttbar (jets) & 540 $\pm$ 160 \\
    Single top & 24 $\pm$ 4\\
    $W$+jets & 21 $\pm$ 12\\
    \hline
    Total expected & 580 $\pm$ 160 \\
    \hline
    Fit result & 520 $\pm$ 97 (stat.) $\pm$ 78 (syst.) \\
    \hline
    \multicolumn{2}{|c|}{\qcd} \\
    \hline
    Fit result & 960 $\pm$ 77 (stat.) $\pm$ 74 (syst.) \\
    \hline
    
  \end{tabular}
  \caption{Comparison of the numbers of events from MC expectations
    and from the results of the fit to the data for the three
    templates. The uncertainties on the MC expectations include the
    systematic uncertainties of the selection efficiency described in
    Sect.~\ref{sec:systematics}.  No MC predictions are available for the 
    \qcd contribution.}
  \label{tab:pred}
\end{table}

To extract the number of signal events, predictions from simulation are used to subtract
the backgrounds from $W$+jets and single-top events (9 $\pm$ 5 and 12
$\pm$ 2, respectively) from the fitted number of \tauele events. The number
is then scaled by the expected ratio, $N_{\tau}/({N_{\tau}+N_{e}})$, of
\tauhad and electrons passing the selection in the \ttbar sample. This
ratio is estimated from MC simulation to be 0.78 $\pm$ 0.03 (stat.) $\pm$
0.03 (syst.).  This yields a final number of observed signal events of
$N_{\tau}$ = 194 $\pm$ 18 (stat.) $\pm$ 11 (syst.).

The cross section is obtained using $\sigma_{t\bar{t}} =
{N_{\tau}}/({{\cal L} \cdot \varepsilon})$. The efficiency
($\varepsilon$) is estimated from MC simulation to be (6.0 $\pm$ 1.4)
$\times 10^{-4}$.  It includes the branching fractions for the various
\ttbar decays and the acceptance, and assumes Br(\ttbar$\rightarrow$\tauhad
+ jets) to be 0.098 $\pm$ 0.002~\cite{PDG}. The efficiency is corrected
for a 13\% difference between MC simulation and data in the trigger
and $b$-tagging efficiencies~\cite{Btagging}. The method used for obtaining
the uncertainty on the cross section is detailed in
Sect.~\ref{sec:xssystematic}.

The cross section is measured to be $\sigma_{t\bar{t}}$~= 194 $\pm$ 18~(stat.) $\pm$ 46~(syst.) ~pb.

% -*- mode: LaTeX; mode: auto-fill; fill-column: 75 -*-
\section{Systematic uncertainties}
\label{sec:systematics}
\label{sec:fitsys}

A summary of all systematic uncertainties on the cross section is given in
Table \ref{tab:uncertainties}.

\label{sec:xssystematic}
\label{sec:efficiency}

\begin{table}[hbtp]
  \centering
  \begin{tabular}{|l|c|}
\hline
Source of uncertainty & Relative uncertainty \\
\hline
ISR/FSR & 15\% \\
Event generator & 11\% \\
Hadronisation model & 6\% \\
PDFs &  2\% \\
Pile-up & 1\% \\
\hline
$b$-jet tagging efficiency & 9\% \\
Jet energy scale & 5\% \\
\met significance mismodelling & 5\% \\
$b$-jet trigger efficiency & 3\% \\  % 2.5
Jet energy resolution & 2\% \\
\hline
Fit systematic uncertainties & 4\% \\
\hline
Luminosity & 4\% \\
\hline
\hline
Total uncertainty & 24\% \\

\hline
  \end{tabular}
  \caption{
    Systematic uncertainties on the \ttbar cross section.
  }
  \label{tab:uncertainties}
\end{table}

The uncertainty on the selection efficiency due to the choice of the
configuration for the MC simulation is estimated by using alternative MC samples and
reweighting procedures. The difference in the efficiency obtained from
various configurations is taken as the uncertainty.  The uncertainty on
the modeling of the ISR/FSR is taken into account by using {\sc AcerMC}~\cite{AcerMC} 
samples with specific tunes aimed at conservatively varying the \linebreak amount of
parton showering~\cite{CSCbook}.  The uncertainty due to the choice of the matrix element event
generator is estimated by comparing \ttbar samples generated using
\linebreak {\sc
  MC@NLO}, {\sc Powheg}~\cite{Powheg1,Powheg2,Powheg3}, and {\sc
  Alpgen}~\cite{Alpgen}. To study the impact of different hadronization
models, events generated using {\sc Powheg} are processed with two
different hadronization programs: {\sc Herwig} and {\sc Pythia}. The
uncertainty of the choice of PDFs is estimated using a number of current
PDF sets~\cite{PDF}.

Uncertainties on the simulation of the detector response are taken into
account using dedicated studies of the reconstructed physics objects
(electrons, muons, jets, \met{}). The uncertainties considered are
associated with the jet energy scale, jet energy resolution, $b$-tagging
efficiency, trigger efficiency and the \met calculation~\cite{jetsys,Btagging}. The uncertainty due to
mismodelling of the lepton veto is estimated using
the uncertainties on the muon and electron reconstruction efficiencies
determined from independent data samples, and is found to be
negligible. 

To obtain the uncertainty on the fit results, variations are applied to the
templates to describe various systematic effects. As the \tauele template
is taken directly from MC-simulated \ttbar events, the systematic uncertainties on this
template are taken from estimates of the mismodelling of the simulation. The
dominant contributions come from variations in the amount of ISR/FSR in the
simulation (1\%), the modelling of the pile-up (1\%), and the
statistical uncertainties (1\%). Uncertainties on the track reconstruction
efficiency, jet energy scale, and the ratio of \tauhad to electrons are found
to be negligible. The \comb template is obtained from a $\mu$ + jets
control sample of \ttbar events in data. The dominant contributions to the
uncertainty come from the
statistical uncertainties (4\%), the difference in shape between the
$\mu$ + jets template and the expected \comb distribution, estimated from MC
samples (2\%), and the MC-based subtraction of the dilepton contribution
(1\%). The uncertainty on the MC-based kinematic correction is found to be
negligible.  The \qcd template is derived from a background-dominated
sideband region with small values of \metsig. The two sources of
uncertainties are the dependence of the template on the \metsig criterion of
the control region, obtained by varying the \metsig requirement (1\%), and
the statistical uncertainty of the control region (1\%).   The total
systematic uncertainty on the fit is found to be 4\%. 

The uncertainty on the luminosity is calculated to be 3.7\% as described
in Ref.~\cite{ATLAS-CONF-2011-116}. The total systematic uncertainty on the cross section is 24\%.

% -*- mode: LaTeX; mode: auto-fill; fill-column: 75 -*-

\section{Conclusions}
\label{sec:conclusions}

This letter presents a measurement of the top quark pair production cross
section in the final state corresponding to the $t\bar{t} \rightarrow
[b\tauhad \nu_{\tau}][bqq]$ decay.   The measurement uses a dataset
corresponding to an integrated luminosity of 1.67~\ifb~of proton--proton
collision data at a centre-of-mass energy of 7 \tev~recorded by the ATLAS
experiment at the LHC. The signal has been extracted by fitting the number of tracks
associated with tau lepton candidates using templates derived from simulation
for the $t\bar{t}$ signal and from the data for the backgrounds.

The \ttbar production cross section is measured to be $\sigma_{t\bar{t}}$~=
194 $\pm$ 18 (stat.) $\pm$ 46 (syst.)~pb. This result is compatible with
the highest precision ATLAS measurements~\cite{Dilep,Slep}, and with the result of 186 $\pm$
13 (stat.) $\pm$ 20 (syst.) $\pm$ 7 (lum.) pb{} obtained in the
complementary tau + lepton (electron or muon) channel~\cite{taulep}.  It is
also in good agreement with the theoretical prediction of
$167_{-18}^{+17}$~pb{}.

% Acknowledgements for papers with collision data
% Version 5-Nov-2012

\section{Acknowledgements}

% Standard acknowledgements start here
%----------------------------------------------
We thank CERN for the very successful operation of the LHC, as well as the
support staff from our institutions without whom ATLAS could not be
operated efficiently.

We acknowledge the support of ANPCyT, Argentina; YerPhI, Armenia; ARC,
Australia; BMWF and FWF, Austria; ANAS, Azerbaijan; SSTC, Belarus; CNPq and FAPESP,
Brazil; NSERC, NRC and CFI, Canada; CERN; CONICYT, Chile; CAS, MOST and NSFC,
China; COLCIENCIAS, Colombia; MSMT CR, MPO CR and VSC CR, Czech Republic;
DNRF, DNSRC and Lundbeck Foundation, Denmark; EPLANET, ERC and NSRF, European Union;
IN2P3-CNRS, CEA-DSM/IRFU, France; GNSF, Georgia; BMBF, DFG, HGF, MPG and AvH
Foundation, Germany; GSRT and NSRF, Greece; ISF, MINERVA, GIF, DIP and Benoziyo Center,
Israel; INFN, Italy; MEXT and JSPS, Japan; CNRST, Morocco; FOM and NWO,
Netherlands; BRF and RCN, Norway; MNiSW, Poland; GRICES and FCT, Portugal; MERYS
(MECTS), Romania; MES of Russia and ROSATOM, Russian Federation; JINR; MSTD,
Serbia; MSSR, Slovakia; ARRS and MVZT, Slovenia; DST/NRF, South Africa;
MICINN, Spain; SRC and Wallenberg Foundation, Sweden; SER, SNSF and Cantons of
Bern and Geneva, Switzerland; NSC, Taiwan; TAEK, Turkey; STFC, the Royal
Society and Leverhulme Trust, United Kingdom; DOE and NSF, United States of
America.

The crucial computing support from all WLCG partners is acknowledged
gratefully, in particular from CERN and the ATLAS Tier-1 facilities at
TRIUMF (Canada), NDGF (Denmark, Norway, Sweden), CC-IN2P3 (France),
KIT/GridKA (Germany), INFN-CNAF (Italy), NL-T1 (Netherlands), PIC (Spain),
ASGC (Taiwan), RAL (UK) and BNL (USA) and in the Tier-2 facilities
worldwide.
%----------------------------------------------

\bibliographystyle{spphys}       % APS-like style for physics

\bibliography{ttbartaujets}

\onecolumn
\clearpage 
% ATLAS Collaboration author list
% Data extracted on 30-Nov-2012 for paper reference TOPQ-2012-04
%\documentclass[11pt]{article}
%\usepackage{a4wide}\begin{document}
\begin{flushleft}
{\Large The ATLAS Collaboration}

\bigskip

G.~Aad$^{\rm 48}$,
T.~Abajyan$^{\rm 21}$,
B.~Abbott$^{\rm 111}$,
J.~Abdallah$^{\rm 12}$,
S.~Abdel~Khalek$^{\rm 115}$,
A.A.~Abdelalim$^{\rm 49}$,
O.~Abdinov$^{\rm 11}$,
R.~Aben$^{\rm 105}$,
B.~Abi$^{\rm 112}$,
M.~Abolins$^{\rm 88}$
% ...
\bigskip
\\
% ...
$^{11}$ Institute of Physics, Azerbaijan Academy of Sciences, Baku, Azerbaijan\\
$^{12}$ Institut de F{\'\i}sica d'Altes Energies and Departament de F{\'\i}sica de la Universitat Aut{\`o}noma de Barcelona and ICREA, Barcelona, Spain\\
% ...
$^{21}$ Physikalisches Institut, University of Bonn, Bonn, Germany\\
% ...
$^{48}$ Fakult{\"a}t f{\"u}r Mathematik und Physik, Albert-Ludwigs-Universit{\"a}t, Freiburg, Germany\\
$^{49}$ Section de Physique, Universit{\'e} de Gen{\`e}ve, Geneva, Switzerland\\
% ...
$^{88}$ Department of Physics and Astronomy, Michigan State University, East Lansing MI, United States of America\\
% ...
$^{105}$ Nikhef National Institute for Subatomic Physics and University of Amsterdam, Amsterdam, Netherlands\\
% ...
$^{111}$ Homer L. Dodge Department of Physics and Astronomy, University of Oklahoma, Norman OK, United States of America\\
$^{112}$ Department of Physics, Oklahoma State University, Stillwater OK, United States of America\\
% ...
$^{115}$ LAL, Universit{\'e} Paris-Sud and CNRS/IN2P3, Orsay, France\\
% ...
\end{flushleft}

%\end{document}
% Created with ./xml2latex.py

\end{document}